\documentclass{article}

\usepackage{arxiv}

\usepackage{amsmath}
\usepackage[utf8]{inputenc} 
\usepackage[T1]{fontenc}    
\usepackage{hyperref}       
\usepackage{url}            
\usepackage{booktabs}       
\usepackage{amsfonts}       
\usepackage{nicefrac}       
\usepackage{microtype}      
\usepackage{lipsum}
\usepackage{cite}
 \usepackage{graphicx}
\usepackage{booktabs}
\usepackage[ruled,vlined]{algorithm2e}

\title{Regression Enrichment Surfaces: a Simple Analysis Technique for Virtual Drug Screening Models}

\author{
  Austin Clyde $\qquad$ Xiaotian Duan $\qquad$  Rick Stevens\\
  Department of Computer Science\\
  University of Chicago\\
  Chicago, IL 60637 \\
  \hfill \\
  Computing, Environment, and Life Sciences Division \\
  Argonne National Laboratory\\
  Lemont, IL 60439 
}

\begin{document}
\maketitle

\begin{abstract}
    We present a new method for understanding the performance of a model in virtual drug screening tasks. While most virtual screening problems present as a mix between ranking and classification, the models are typically trained as regression models presenting a problem requiring either a choice of a cutoff or ranking measure. Our method, regression enrichment surfaces (RES), is based on the goal of virtual screening: to detect as many of the top-performing treatments as possible. We outline history of virtual screening performance measures and the idea behind RES. We offer a python package and details on how to implement and interpret the results. 
\end{abstract}


\section{Introduction}
With the recent surge in deep learning research, metrics for model success have remained relatively stagnant. Data, models, and application domains seem endless, but the methods used to gauge success in the transfer of the model from research space to tools in real-world applications have rested on intrinsic value in mean squared error, classification accuracy, or assumption-based metrics such as $r^2$ score. In the classification side of the field, such as a subset of computer vision tasks, the metrics seem to represent the problem at hand truly. The accuracy of an image labeling system is representative of the real-world application, in classifying images for web searches or automatic keyword generation \cite{krizhevsky2012imagenet, deng2009imagenet}. Domains such as natural language processing (NLP) have recognized this problem as they often employ specialty metrics such as the BLEU score \cite{papineni2002bleu, chen2011amber, post2018call}. In the new space  of applying deep learning for drug discovery and virtual screening, the metrics used  to score  models have not moved towards the necessary specialization for exhibiting confidence and understanding their performance for the use in real lead discovery pipelines \cite{unterthiner2014deep, zhang2017machine, pereira2016boosting, schneider2010virtual}. 

\subsection{Virtual Screening}
 A common problem in the drug discovery process is finding the few active drugs in the trove of the imaginable drug-like compounds, estimated to be around $10^{63}$ \cite{bohacek1996art, fink2005virtual}. Recent literature from medicinal chemistry has shown lingering inductive biases even in vast vendor libraries, indicating the idea of sub-sampling a library for testing as an ineffective means of screening, providing an impetus towards screening hundreds of millions to billions of diverse compounds rather than small curated libraries \cite{jia2019anthropogenic, cleves2008effects}. Virtual screening is a computational technique for identifying possible subsets of hits suitable for downstream analysis with more expensive computational or lab experiments. Virtual screening is typically applied a costly or time-consuming process to select a subset of objects to either run those costly experiments on or pursue downstream work on.  Virtual screens are applied to the biological sciences for drug screening, though recently there have been cross-over from material designs as both fields pursue machine learning screening workflows \cite{halls2013virtual}. For example, virtual structural docking is a technique used to score compounds to understand how well, and if, a compound docks to a particular ligand \cite{lyne2002structure}. Producing an experimental value requires a time consuming experimental process, and is not possible at the scales of compounds that are screened virtually \cite{lyu2019ultra}. With current high throughput laboratory and computational techniques, these sets can be large, even in the hundreds of millions in the case of \textit{Lyu et al.}

In many ways, virtual screening is a ranking problem \cite{kontoyianni2005evaluation, wilton2003comparison, kellenberger2008ranking, hawkins2007comparison}. Given a set of drugs from a vendor library,  researchers want an ordering of which compounds are likely to work where downstream experimentation or computation can begin. While classification metrics are used in the literature as a surrogate for the \textit{what researchers are interested in} question, it is not uncommon in the lead discovery process to not know \textit{a priori} what the cutoff or size of a desired hit set is---that is likely to be a function of budget, lab constraints, etc. By casting the problem in terms of rank ordering, it separates the development of the model and the function of the model into research and development workflows. We want to use a model to rank the compounds in our domain, and we would like that ranking to align well with the actual rankings if finding those rankings were tractable. The field of information retrieval has created a variety of metrics based on document interest retrieval such as expected reciprocal rank (ERR) and discounted cumulative gain (DCG) \cite{chapelle2009expected}
\begin{equation}
    ERR = \sum_{i=1}^n\frac{1}{r}P(\text{User is satisfied with r}) \qquad DCG = \sum_{i=1}^k \frac{2^{g_i}-1}{\log(i+1)}
\end{equation}
where $\psi(r)$ is the utility of a given rank (so $\psi(1)=1$ and $\psi(x)\rightarrow 0$ as $x\rightarrow \infty.$
In reality, the distributions encountered in this area are highly skewed, with a vast majority of compounds failing to work or being \textit{interesting} in any sense to the experimentalist. As the rankings fail from the top two or three, the interest in their relative accuracy diminishes rapidly. Some progress has been made in ranking metrics though they generally fail to capture the heavy tail focus this problem exhibits and are ineffective in data with high regions of error propagating from experimental or random processes \cite{wang2014learning, zou2016novel, balakrishnan2012collaborative, cossock2006subset}. The application of ranking measures to deep learning models is still in its infancy, leading to a disappearance of ranking from very recent literature with deep learning screening models. 

\subsection{Virtual Screening Metrics} 
Before moving on to discuss the difficulties associated with evaluating ML models in a virtual screening context, we outline a history of virtual screening metrics. In these settings, the targets were typically experimentally derived hits or known leads. In this setting, rankings or scores were typically not computed on the targets themselves, only on the predicted or computed values. New measures were created, such as the Boltzmann-Enhanced Discrimination of ROC (BEDROC) scores \cite{truchon2007evaluating}, enrichment factors, and ROC Enrichment (ROCE) \cite{nicholls2008we}, robust initial enrichment (RIE) \cite{sheridan2008multiple}, and sum of the log of ranks test (SLR) \cite{zhao2009statistical}. Borrowing notation from Trunchon, let $N$ be the study size and $n$ be the number of active compounds. We will assume that there is some cutoff $\delta$ so that true hits are labeled as either a hit or not in the case that the true values are provided as real values. Let the cumlative density function $F_a(x)$ be the probability that an active compound is found prior to rank $x$, so if there are 1000 compounds in a screen, $F_a(10)$ is the probability that the top ten molecules from the screen provide a single hit. The probability density function is defined as $f_a(x)$ which is the probability the relative rank $x$ is a hit. Using this CDF function, the area under the accumulation curve (AUAC) is simply equal to \cite{kairys2006screening, truchon2007evaluating}
\begin{equation}
    \text{AUAC}=\int_0^1 F_a(x)\, dx=1-\langle x \rangle
\end{equation}
where $\langle x \rangle$ is the average rank of a hit. ROC can be formulated form AUAC as \cite{truchon2007evaluating}
\begin{equation}
    \text{ROC} = \frac{\text{AUAC}}{R_i}-\frac{R_a}{2R_i}\quad\text{where}\quad R_a=\frac{n}{N}, R_i=1-R_a
\end{equation}

As you can see in these formulations, the metrics cannot be tuned for early versus late recognition (see figure \ref{}). The insight came as data sizes grew and docking programs were screening larger than before libraries, where late recognition became costly. Eventually, moving the study towards early versus later recognition of hits, weighting schemes were created $w(x)$ such as $e^{-\alpha x}$
\begin{equation}
    w\text{RIE}=\frac{\int_0^1 f_a(x)w(x)\, dx}{\int_0^1 w(x)\, dx}\qquad w\text{AUAC} =\frac{\int_0^1 F_a(x)w(x)\, dx}{\int_0^1 w(x)\, dx}
\end{equation}
Notice in these formulations all measures are dependent on $N,n$ and $\alpha$ such that \cite{truchon2007evaluating}
\begin{equation}
\text{RIE}_\text{max}=\frac{1-e^{\alpha R_a}}{R_a(1-e^\alpha)}\qquad \text{RIE}_\text{min}=\frac{1-e^{-\alpha R_a}}{R_a(1-e^{-\alpha})}.
\end{equation}

Relating these formulations by scaling and weighting
\begin{equation}
    \text{BEDROC} = \frac{\text{RIE}-\text{RIE}_\text{min}}{\text{RIE}_\text{max}-\text{RIE}_\text{min}}=\frac{w\text{AUAC}-w\text{AUAC}_\text{min}}{w\text{AUAC}_\text{max}-w\text{AUAC}_\text{min}}.
\end{equation}

The most recent development for analysis is SLR \cite{zhao2009statistical}
\begin{equation}
    \text{SLR}=\sum_{i=1}^n \log(r_i) \quad\text{where}\quad -\sum_{i=1}^n \log(\frac{r_i}{N}) \sim Gamma(n,1).
\end{equation}
The metric and its interpretation is dependent on $n$ and $N$ as well as the distribution of the ranking in the data. 

All of these measures require at least one parameter setting a cutoff for marking a result a hit or not. AUROC represents the probability of active compounds ranking earlier than decoy or inert compounds. The BEDROC is related to robust initial enhancement (RIE) and both of them represent tuning the AUROC focus from late recognition to early recognition \cite{zhao2009statistical}. In the framework from \cite{nicholls2008we}, these measures are satisfactory as they require parameter choices and change based on the data. Given the importance of cost structure analysis moving forward with deep leaning models in real world applications, we would argue they are also not interpretable in the desired sense. 

Enrichment factor (EF) is most commonly used and is similar to a basic notion of hit rate (HR) \cite{kruger2010comparison, bender2005discussion, stahl2000modifications, shoichet2004virtual, pereira2016boosting, ericksen2017machine, ain2015machine} where $\chi$ is used to represent a hit typically in terms of percentages so $\chi=0.25$ means everything in the top 25\% of the distribution is labeled a hit
\begin{equation}
\text{EF} = \frac{\text{Hits}_\text{sampled} / N_\text{sampled}}{\text{Hits}_\text{sampled} / N_\text{total}}=\frac{\int_0^\chi f_a(x)\, dx}{\chi},
\end{equation}
and gained popularity in the literature for comparing various docking methods. In this method, multiple cutoffs are used to translate docking scores, a continuous measure, to a hit classification. Yield hit rate (YHR) is another commonly seen metric which is effectively the same as EF in the literature \cite{harper2001prediction, ghosh2006structure}

\subsection{Towards Deep Learning Surrogate Screening}
In the relatively new area of virtual screening with statistical and machine learning (ML) techniques, machine learning models are used to replace docking programs or introduce early funnel stage lead discovery techniques \cite{chen2018rise}. Metrics for ML models are typically the only measures technicians have for answering questions regarding model convergence, parameter tuning, and learning success. In a sense, a metric for communicating the  ML model utility must both satisfy the requirements for virtual screening as well as requirements for success in the ML community. While the analysis for most virtual screening metrics rely on a predetermined notion of hit, the sheer size and new domains ML models are creating has lead to a fuzzier notion of hit, where  both the values being predicting on and the predictions themselves are continuous representations of a property (such as cell growth, docking scores, binding affinity $\delta G$ values, etc.). 

In the literature, most authors still report $r^2$ scores or area under the receiver operating characteristic curve (AUROC) with a cutoff \cite{feinberg2018potentialnet, unterthiner2014deep, korotcov2017comparison, hamanaka2017cgbvs, wallach2015atomnet, gonczarek2016learning, menden2013machine}. These metrics are standard to present in the ML community as a means of indicating a model matches a notion of being trained.  Prior to the introduction of ML, metrics in the field were often used to compare docking methods, not to both evaluate the training a model as well as the usefulness of a model---two distinct notions, as structural docking did not have a notion of model training or accuracy in the same way ML models do. At first, the choice of presenting AUROC for deep learning model performance was a sensible choice as virtual screening with structural docking has a large history of AUROC as a means of comparing scoring functions, programs, and efficacy \cite{triballeau2005virtual, nicholls2008we}. ROC curves for virtual screening were preferred as they allowed for regression or uncertainty in a classification sense, though the targets or known hits had to be binned. The ROC curve plots sensitivity and specificity representing various cutoffs for marking a screened value a hit or not. Overtime, the curves would be omitted and represented solely by the AUROC, where the single value provides little information regarding early or late recognition problem and represents a balanced accuracy measure than the early ranking problem truly desired.  

Of course, reporting these scores provides useful information regarding a general overview of the model performance (i.e. whether the model converged, is over-fitting, \textit{working} in a usual sense); yet, $r^2$ or MSE does not provide actionable insight into how well this model will screen new compounds or treatments. If a drug company wanted to understand the cost-benefit analysis for lead development, a $r^2$ score does not provide insight into that computation. The choice of cutoff is also highly dependent on the downstream task, thus radically changing the actual scoring and recognition qualities of a model. Budget choices or downstream experiments are typically made by new progress in models and technology, but the current scoring techniques used for these models requires parameters based on the circular requirements. 

There has been progress on converting these methods for regression models using some normality assumptions such as regression enrichment surface (REF) \cite{feinberg2018potentialnet}
\begin{equation}
\text{EF}_\chi^{(R)} = \frac{1}{\chi\cdot N}\sum_{i}^{\chi\cdot N}\frac{y_i-\overline{y}}{\sigma(y)}
\end{equation}
where the experimental values $y_i$ are ranked according to the model $\hat{y_i}$. This method is based on the distribution of the underlying data as it is normalized by $\sigma(y).$ This method should be preferred in regression settings; however the value itself is still dependent on some $\chi\%$ cutoff decreasing the interpretability across settings.


\section{RES Analysis}

Regression enrichment surfaces is a method for producing plots and a metric to indicate how the enriching a proposed model is for virtual screening. 

Assume we are given some true property values $\{y_i\}_{i\in I}$ and predicted property values $\{\hat{y}_i\}_{i\in I}$ from a model. We produce rankings for both sets, which can be based on some ordering function such as $<$ where $\hat{R}\subseteq I$ is an ordered list of the rankings for the true values and $\hat{R}\subseteq I$ is an ordered list of the rankings for the predicted values. We use the notation $R_{\chi}$ to denote top fractional rank, so $R_{0.25}$ is the top quarter ranks.  

We say the enrichment at some screening cutoff $\rho$ and some true distribution cutoff $\sigma$ is 
\begin{equation}
\text{Enrichment}(R, \hat{R}, \sigma, \rho)=\frac{|R_{\sigma} \cap \hat{R}_{\rho}|}{|R| \text{min} (\sigma, \rho)}.
\end{equation}
This enrichment score is sampled on a log-scale grid for reasonable values. For instance, if the virtual screening evaluation data has 1,000 values, we would sample from $10^{-3}$ to $0$, while $10^{-4}$ to $0$ would be undefined since there is no relative ranking at $0.01\%$ for $R$ in that case. 

\begin{algorithm}[H]
\SetAlgoLined
  \For{$\rho$ in LogScale(0, m)}{
    \For{$\sigma$ in LogScale(0, m)}{
        grid[$\sigma$][$\rho$] = $\text{Enrichment}(R, \hat{R}, \sigma, \rho)$ \;
    }}
 \caption{Regression Enrichment Surface}
\end{algorithm}

The grid of enrichment's is then used as input to a contour plot, for example, where the performance can be visualized. The color represents the accuracy of the screen at given cutoff pair. Further, for a comparison of model performance, the only requirement is that all models compared have the same cutoff bounds $m$. From there, the 2D integral can be taken 
$$ \text{RES Score} = \int_{0}^{1}\int_0^{1} \text{Enrichment}(R,\hat{R}, \sigma, \rho)\,d\sigma\, d\rho.$$ 
While the RES Score can be computed, the chart itself is the most beneficial piece of analysis from the method. 

 The code is available on GitHub and PyPi as a python package \url{https://github.com/aclyde11/regression_enrichment_surface}.

\section{Discussion}
We have presented a new technique for analyzing and interpreting virtual screening models, especially in the context of deep learning model performance analysis. In a sense, the technique is an extension of enrichment factor analysis for multiple parameters; however, we view this extension as essential to understanding at what hit cutoffs would a model be useful. For example, it may be the case that decreasing the cutoff for a certain model screen sensitivity may in fact increase the number of hits delivered to the consumer of the model.  

To provide clarity at how this technique provides an in-depth insight into model performance, we will work through an example workflow. The National Cancer Institute screens thousands of cancer drugs in order to find cures to rare understudied cancers \cite{williams2013patient, reinhold2014nci, belizario2016using}. Their goal is to find new compounds from a corpus that will provide reasonable leads for the next stage of the study, typically patient-derived xenograft (PDX) models. One method is to use high throughput laboratory techniques to screen 60 standard cell lines with compounds. This method is highly efficient as the NCI has screened over 50k compounds, but current virtual libraries contain billions of drug-like molecules \cite{shoemaker2006nci60}. One deep learning application could be to replace the cell line screens with a deep learning model capable of performing inferences on drugs at a rate of hundreds per second \cite{chen2016gene}. 

Suppose we created a few deep learning models $A$, $B$, $C$, and $D$. We want to show the efficacy of these models as a surrogate to \textit{in-vitro} cell line experiments. We are not trying to show the deep learning models are better at predicting hits that concur with PDX experiments (that is an assumption left to the biological sciences). We will use this analysis to show that using virtual screening is on-par with cell line screening or an excellent way to narrow a set down for cell-line screening. Typically metrics such as $r^2$ or MSE from table \ref{tab:performance} are used to gauge the relative performance of the models. It is clear that model $C$ and $D$ seem to correlate in a much stronger sense, and the $\text{EF}_\chi^{(R)}$ scores show some differences, however, with the scaling issue, the performance in a global sense remains unclear. While the RES shows a correlation with the other metrics, the number alone is not the most useful aspect.

\begin{table}[h]
\centering
\label{tab:performance}
\begin{tabular}{@{}lllllllll@{}}
      & \multicolumn{4}{l}{Regression Metrics}                 &  & \multicolumn{3}{l}{Classification Metrics} \\ \cmidrule(lr){2-5} \cmidrule(l){7-9} 
Model & $r^2$ & MSE & EF$_\chi ^{(R)}$ & RES &  & Accuracy & AUC & EF        \\ \midrule
$A$     & 0.76 & 0.0034 & 2.12 &  0.8412 &  &\textbf{ 98.9\%} & 0\textbf{.709}  &  1.0       \\
$B$      & 0.74 & 0.0037 & 2.12 &  0.8421 &  & 98.8\% & 0.647 & 1.0           \\
$C$     & \textbf{0.82} & \textbf{0.0026} & \textbf{2.16} &  \textbf{0.8586} &  & \textbf{98.9\%} & 0.678 & 1.0           \\
$D$    & \textbf{0.82} & 0.0028 & 2.15 &  0.8470 &   & \textbf{98.9\%} & 0.667 & 1.0            \\ 
\bottomrule
\end{tabular}
\caption{Performance metrics across models. $EF_\chi$ is computed at $10\%$. RES bound is set to $10^{-4}$ as the sample size of 50k. Classification metrics are binned at 0.5, which splits the distribution into $2\%$ hits and $98\%$ none-hits. EF is computed at $10\%$ in addition to cutoff. Based on the above scores alone, it seems models A, C, and D are all comparable with $C$ being the best choice.}
\end{table}

In figure 1, we show the four associated RES plots. Recall, a point on the plot indicates how the top $x\%$ captures the top true or $y\%$ (normalized such that the model is not required to capture more than the subset size). There are a few immediate interpretations to note. First, we can read the darkest regions where the percentage is 100\% as relatively safe zones for using these models. All models with the exception of $C$ would capture 100\% of the top drugs if we performed our downstream work on the top 10\% from the model predictions---this is an order of magnitude improvement over not using a virtual screen, and RES method indicates this immediately while measures alone do not. If this were the desired use case of the model, the scores alone from table \ref{tab:performance} would mislead one to use model $C$ when in fact, model $C$ performed the worst on this task. RES plots do indicate that model $C$ would be the best choice of model if the downstream task was more selective and required selectivity of only $10^{-3}$ candidates for example, in which case the model seems to detect the very top much better than the other models (you can see how the contours curve out at the bottom of model $C$ around $10^{-3}$ on the $x$-axis while the others are relatively straighter. 

In the case of comparing model performance, one can take the difference of the plots over various models (Figure \ref{fig:difference}). In this setting, we see that model C performs much better at very high relative ranks around $10^{-2}$, but we can also see that at higher ranges $10^{-1}$, model B might actually have better detection in higher ranges. This view also opens an interesting research question regarding the landscape of predictions that result from the use of different feature modes.

\begin{figure}[h]
    \centering
    \includegraphics[width=0.45\textwidth]{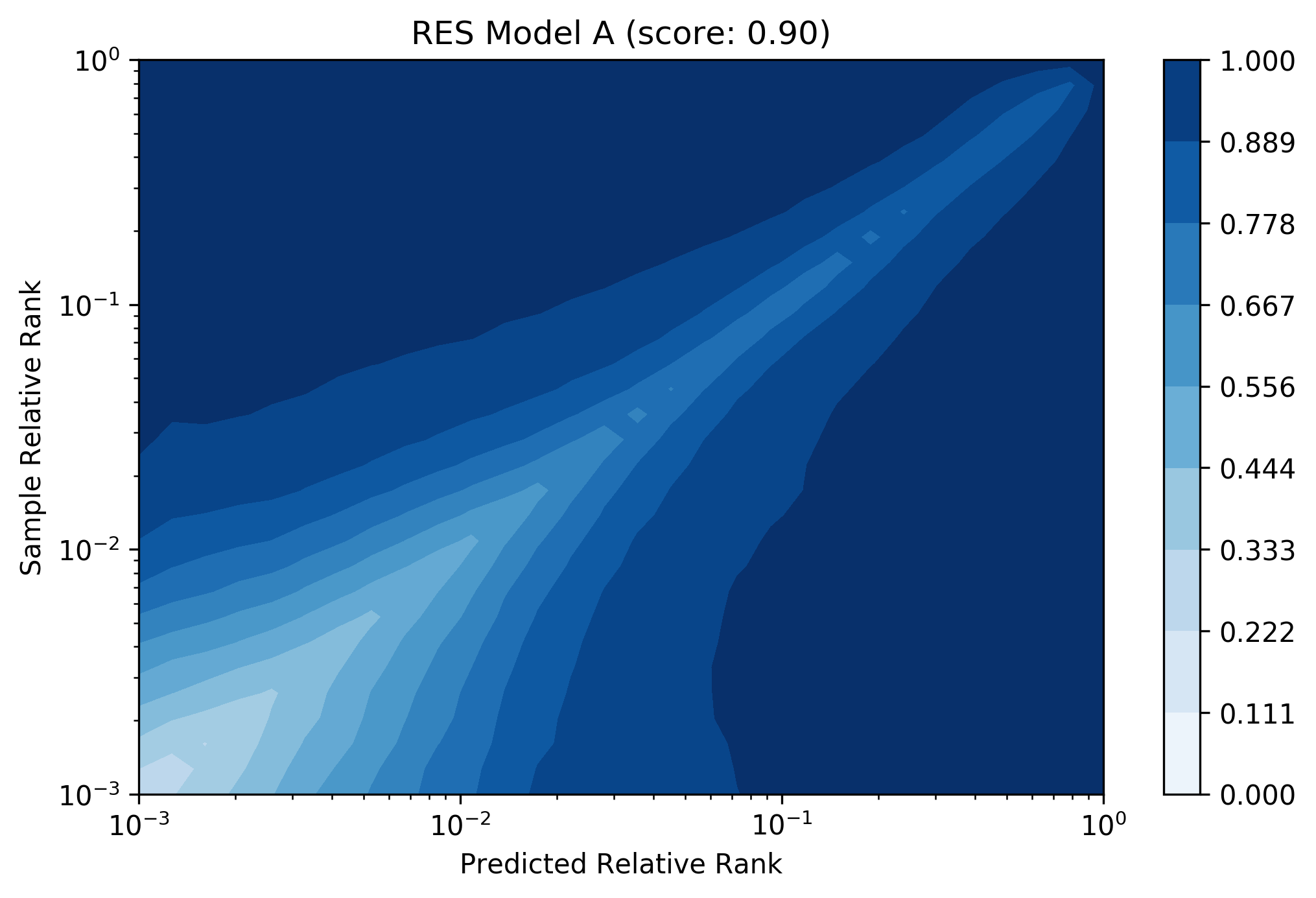}
    \includegraphics[width=0.45\textwidth]{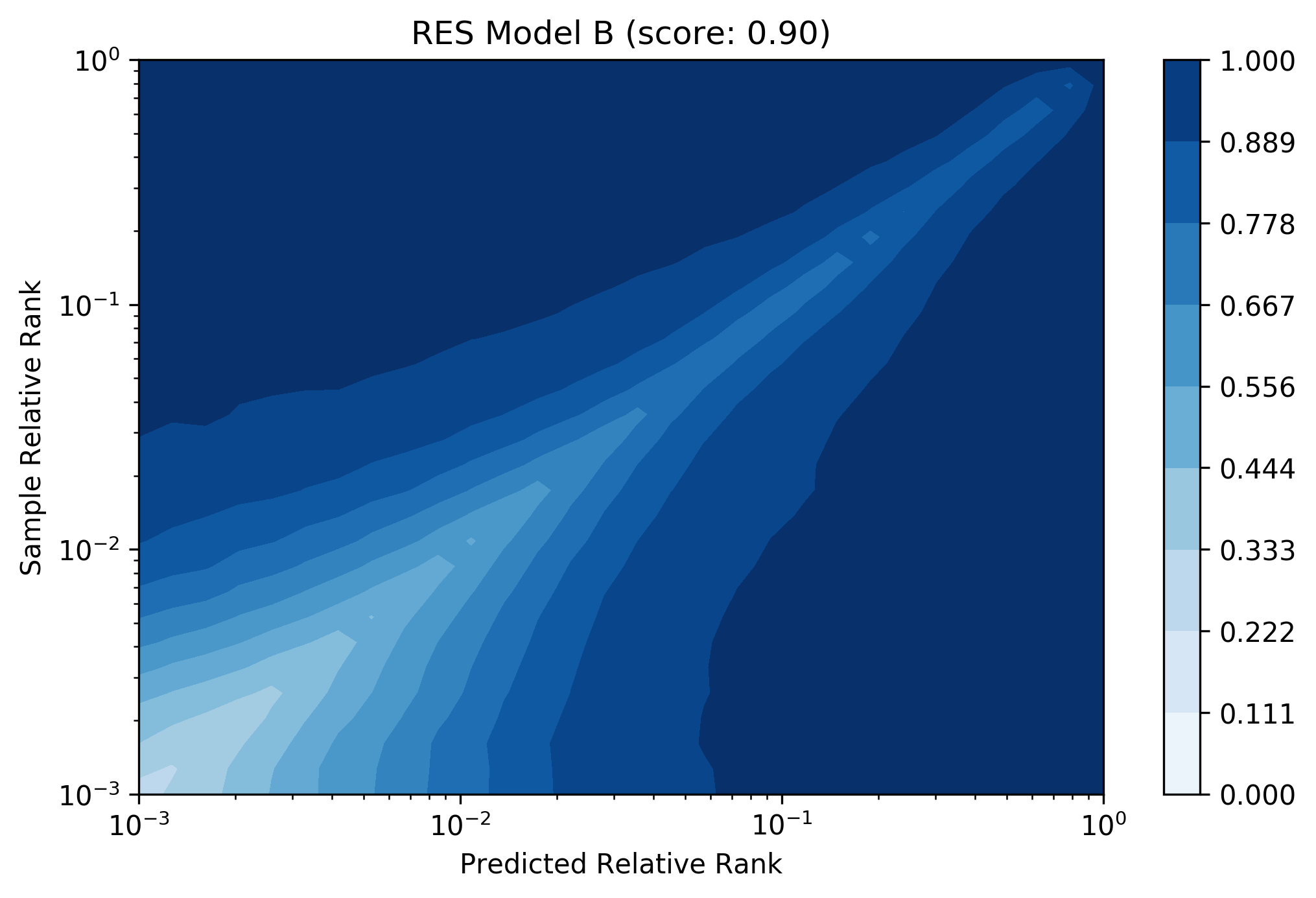}
    \includegraphics[width=0.45\textwidth]{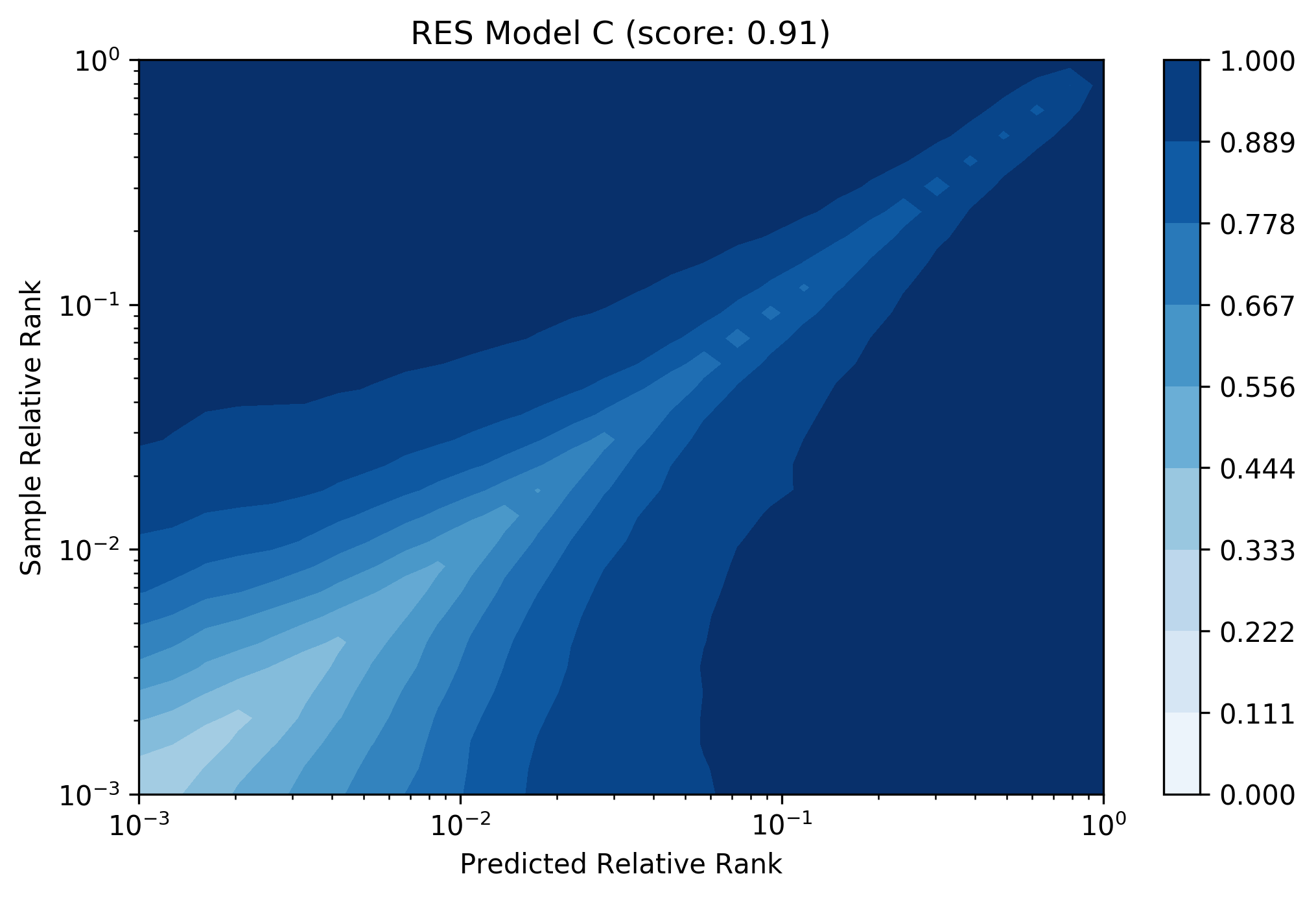}
    \includegraphics[width=0.45\textwidth]{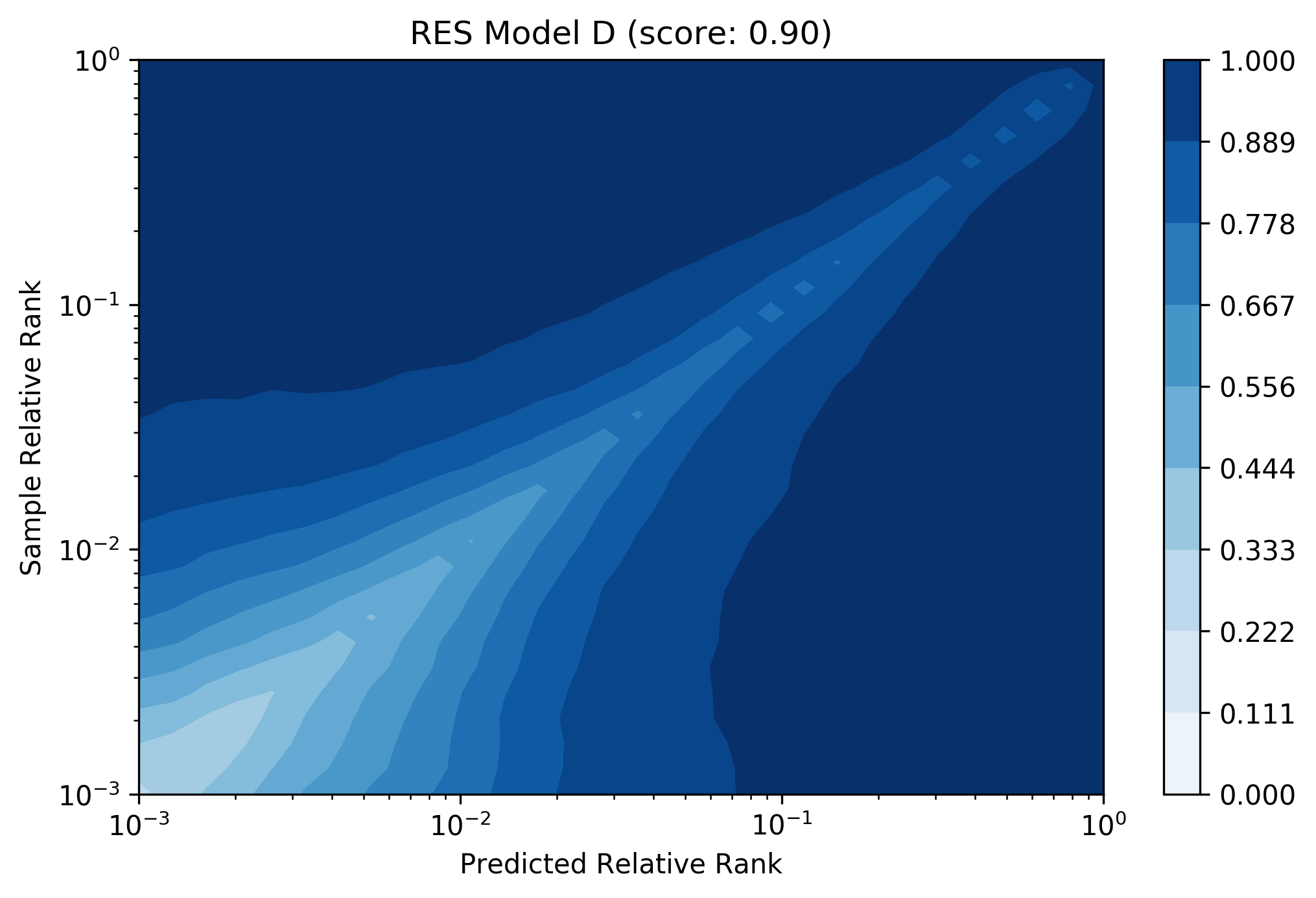}
    \caption{Regression enrichment surface (RES) plots for the associated model and predictions. The RES score noted in the title is an approximation of the integral where the bounds are alerted to be 0-1 for both $x$ and $y$ axis such that the best performance is 1 and the worsts performance is 0. It should be noted that the original bounds should be communicated so that the score can correctly be reported and reproduced.}
    \label{fig:res}
\end{figure}

\begin{figure}
    \centering
    \includegraphics[width=0.3\textwidth]{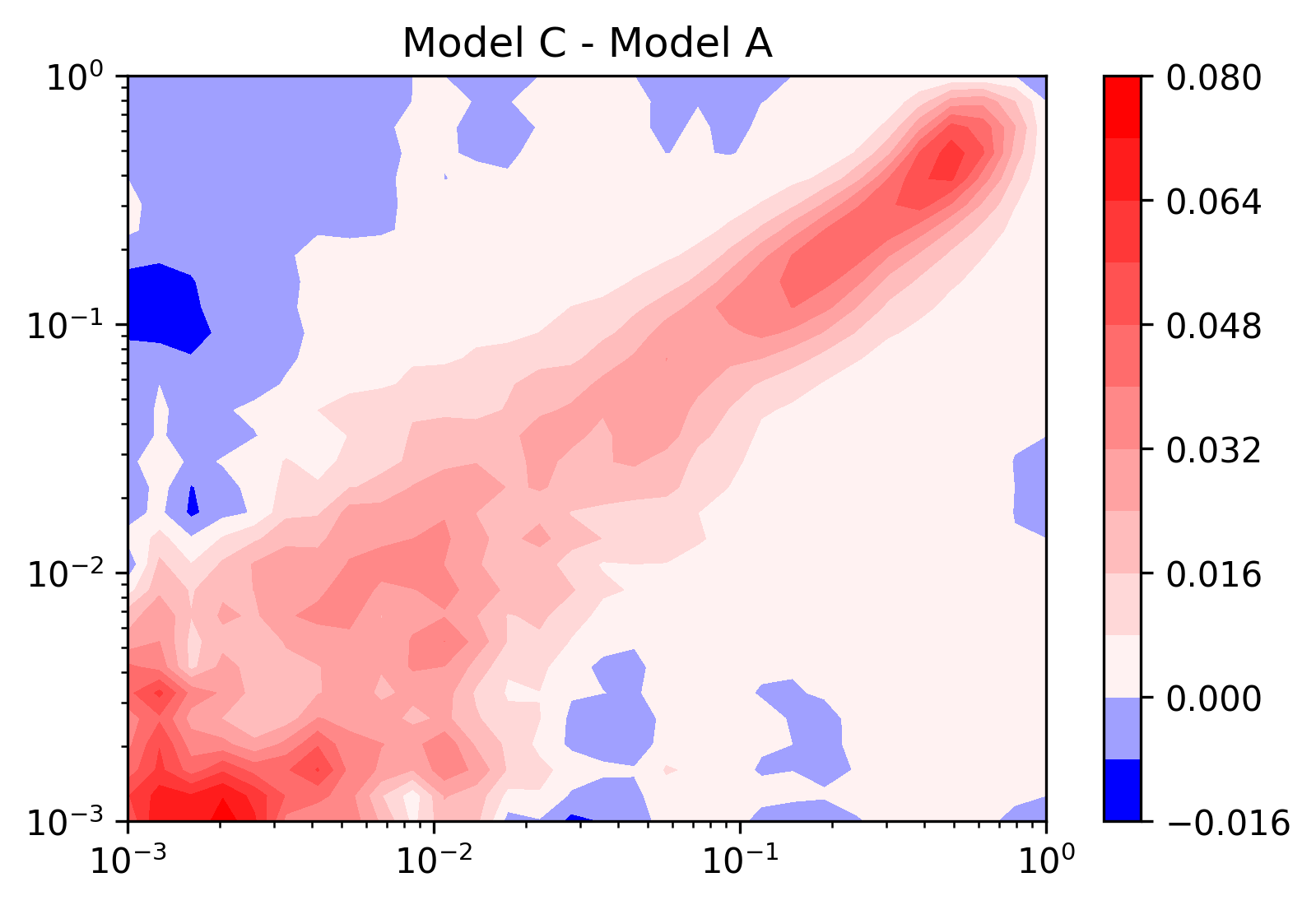}
    \includegraphics[width=0.3\textwidth]{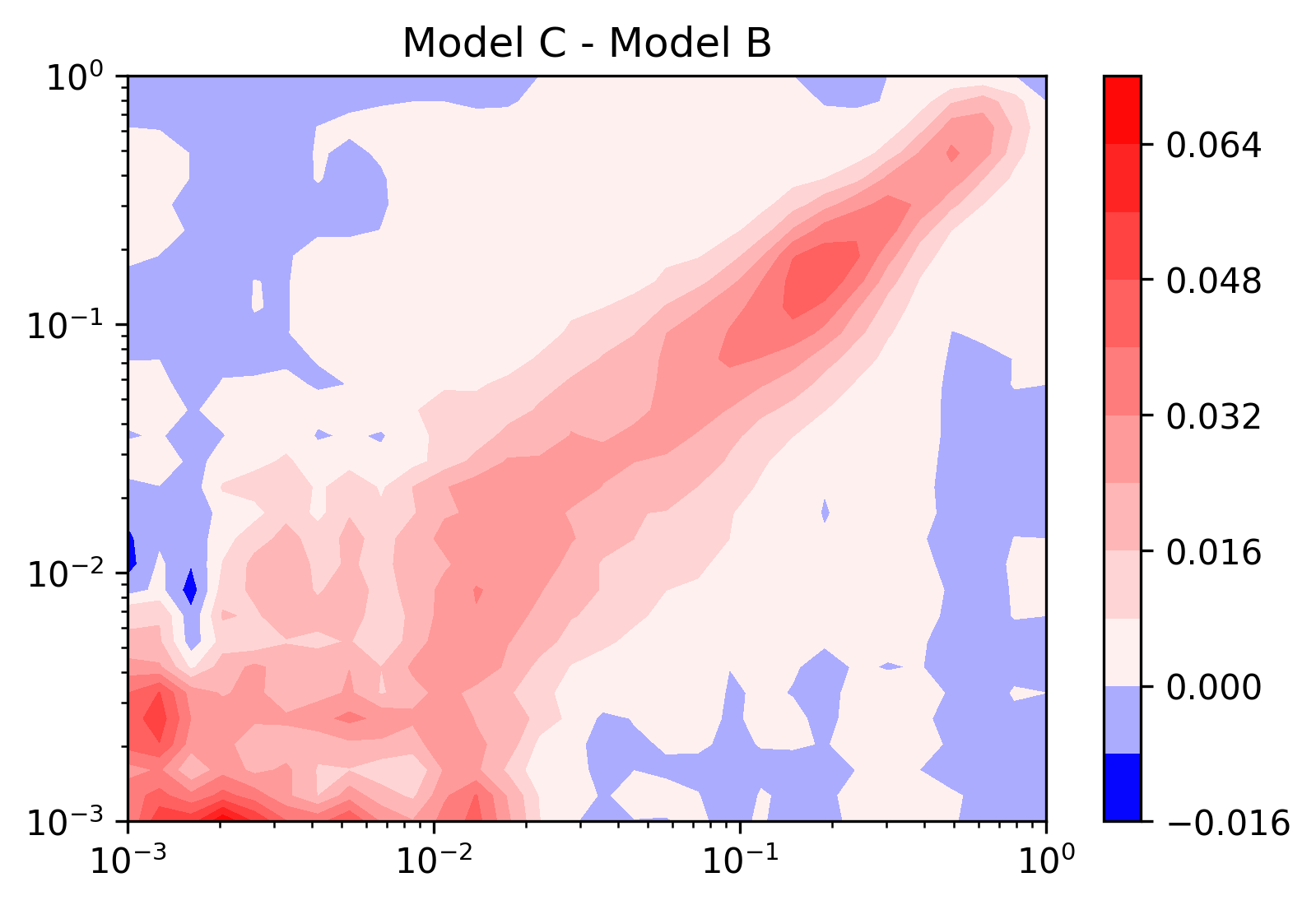}
    \includegraphics[width=0.3\textwidth]{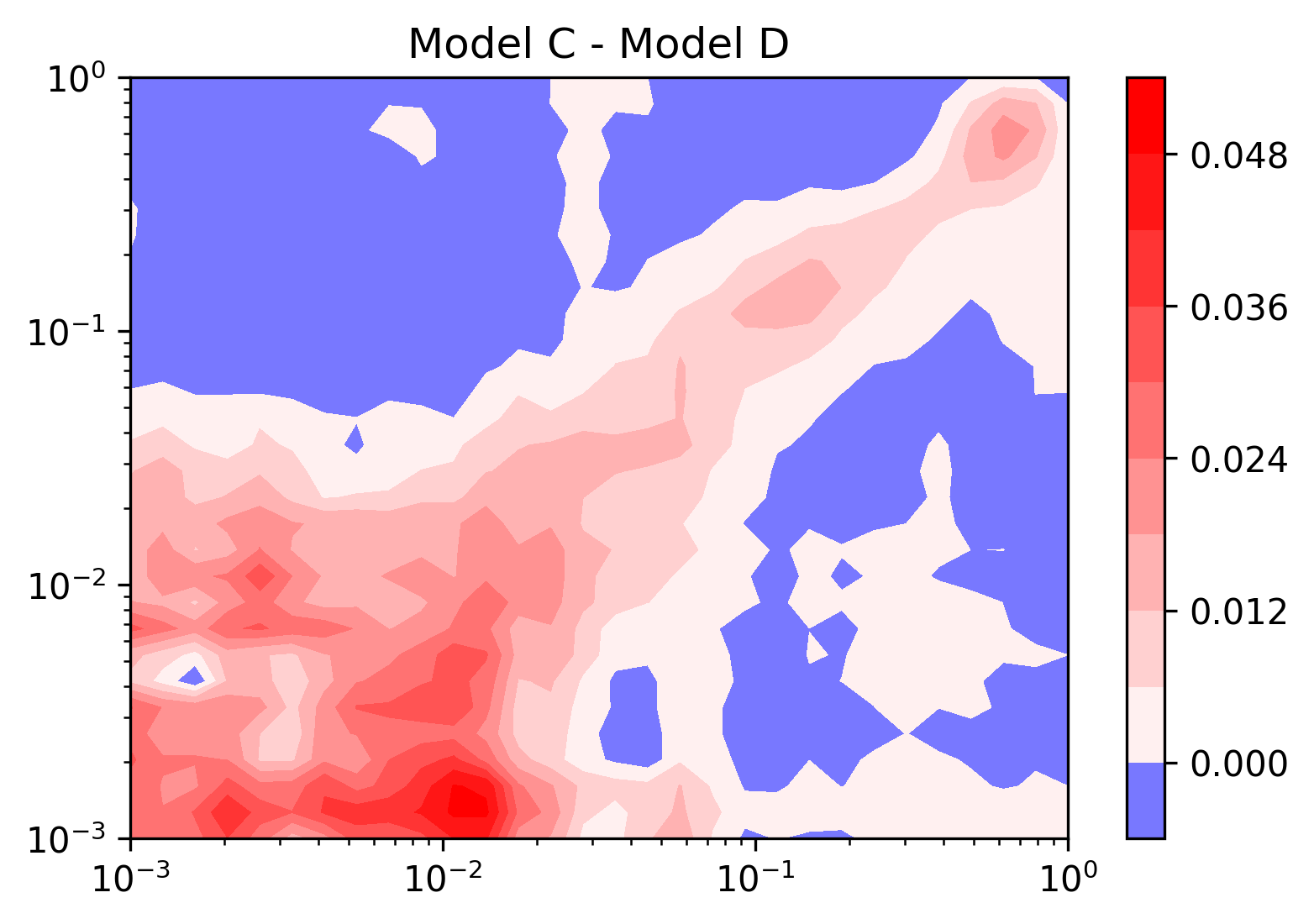}

    \caption{The difference was taken between the enrichment surface plots of models A, B, D with respect to C. From this view, it is clear to see in what regions and tasks model C may perform better.}
    \label{fig:difference}
\end{figure}

\section{Conclusion}
RES analysis captures the application-use case of deep learning models for virtual screening. While other common statistical metrics are useful in the model training process, RES plots indicate performance in terms of the downstream task requirements (cost, throughput, etc).

 \section*{Acknowledgement}
This research used resources of the Argonne Leadership Computing Facility, which is a DOE Office of Science User Facility supported under Contract DE-AC02-06CH11357.

\bibliographystyle{unsrt}  
\bibliography{references}

\end{document}